\documentclass[fleqn,10pt]{wlscirep}
\usepackage[utf8]{inputenc}
\usepackage[T1]{fontenc}
\usepackage{siunitx}
\usepackage{xcolor}
\usepackage{textcomp}
\usepackage[normalem]{ulem}
\title{Complementary and Asymmetric Tapered Bent Mid-Infrared Waveguide Arrays for Subwavelength-Pitch Integration and Crosstalk Minimization }

\author[1]{Humaira Zafar}
\author[1,*]{M. F. Pereira}

\affil[1]{Department of Physics, Khalifa University, Abu Dhabi 127788 UAE}

\affil[*]{mauro.pereira@ku.ac.ae}


\begin{abstract}
This paper delivers the first report of a mid-infrared (MIR) waveguide array design that employs complementary and asymmetric tapered Euler-shaped bends. These provide greater fabrication flexibility to achieve subwavelength-pitch integration while reducing crosstalk to below 30 dB across the \SI{3.1}{\micro\meter} to \SI{3.6}{\micro\meter} wavelength range. Unlike previous designs, which maintained constant waveguide widths, the Euler waveguide bends are characterized by asymmetric and complementary tapered waveguide widths. This approach significantly reduces crosstalk to below 30 dB for both the first and second neighboring waveguides across a 500 nm wavelength range, enhancing the efficiency of optical phased arrays (OPA) with a large field of view, optimizing light propagation and minimizing crosstalk. The waveguide array is fabricated on a silicon-on-insulator platform, with a 2-micron buried oxide layer and a 500 nm-thick silicon layer. The design is highly tolerant to fabrication variations, maintaining consistent performance even with width variations. The spectral responses, simulated using the 3D finite-difference time-domain method, demonstrate negligible coupling and low insertion loss across the wavelength range. This work offers a robust and CMOS-compatible solution for MIR integrated photonic circuits.

\end{abstract}
\begin{document}

\flushbottom
\maketitle
%
%
\section*{Introduction}

Mid-infrared (MIR) photonics and optoelectronics can exploit the unique absorption characteristics of molecules in this spectral range to create compact and low-power sensors for diverse applications, including healthcare, environmental monitoring, security, industry, transport, agriculture and free space communications \cite{Grillot:2025,Scalari:2019, Karabchevsky2,shekhar2024roadmapping, apostolakis2024photoacoustic,Cousin2022}. The main MIR source, the quantum cascade laser (QCL) \cite{Faist:1994,Vitiello:15}, can now be grown on silicon substrates \cite{CorbettNature2025,nguyen2018quantum}, indicating the possibility of full on-chip integration of source and photonic functionalities\cite{Karabchevsky1}. Additionally, this technology can greatly minimize the size of existing astronomy setups by miniaturizing optical systems into photonic chips and enable free-space communications like light detection and ranging (LiDAR), (transmission windows of \SIrange{3}{5}{\micro\meter} and \SIrange{8}{12}{\micro\meter}) \cite{spitz2021free} highlighting the necessity for developing MIR devices in silicon photonics and the timeliness of this application.  

Furthermore, it is an emerging platform for photonics integrated circuits (PIC) due to its compatibility with CMOS technology. Although much research has focused on the near-infrared (NIR) range, silicon's transparency up to \SI{8}{\micro\meter} has led to the proposal of several MIR modulators, detectors, couplers, splitters, polarizers, and resonators \cite{review,ZafarIEEEAccess,nedeljkovic2013silicon}. Dense waveguide (WG) arrays with subwavelength pitches are essential for applications in wide field-of-view optical phased arrays (OPA) used for beamforming and steering in LiDAR systems (OPAs can replace mechanical beam steering in LiDAR, enabling faster and more reliable scanning). As the first two neighboring waveguides at the subwavelength pitches on either side of the input waveguide experience light transfer, a significant crosstalk reduction is necessary for efficient OPA and LiDAR systems. 
There are numerous proposals for dense waveguide arrays in the NIR, however, only one report exists in the MIR \cite{li2024morphology}, targeting the wavelength range of \SIrange{3.5}{3.6}{\micro\meter}, where the authors employ Bézier curves to reduce crosstalk, achieving a reduction to 30 dB in both neighboring waveguides over an approximately \SI{40}{\nano\meter} wavelength range.

Dense waveguide arrays require effective suppression of light coupling between neighboring waveguides. Reducing crosstalk is particularly challenging, especially for waveguide arrays with subwavelength pitches. Achieving subwavelength pitch waveguide arrays with minimal crosstalk is crucial to ensuring the efficiency of optical phase arrays. Waveguide arrays employing various approaches have been reported in the literature for the NIR spectrum. These include waveguide superlattices with different widths of the neighboring waveguides \cite{song2015high}, inverse design \cite{shen2016increasing}, subwavelength silicon strips \cite{chen2022realization}, and bent waveguide arrays with various shapes \cite{zafar2023band, zafar2023compact, xu2016ultra}. The modulation of the bent shapes of the waveguides, also referred to as artificial gauge fields, has led to the development of new coupling strategies aimed at minimizing crosstalk at small pitches \cite{li2024morphology, zhou2023artificial, lumer2019light, song2022dispersionless, yi2020design}. All the waveguide arrays mentioned above have been designed for the NIR spectrum, except for one report in Ref. \cite{li2024morphology}. Reference \cite{li2024morphology} describes a waveguide array that uses Bezier curves to reduce crosstalk. In all previously reported bent waveguide arrays, the crosstalk reduction could not go below 30 dB over a broad wavelength range for both the first and second neighboring waveguides. Therefore, further exploration of bent configurations is needed to control the coupling between subwavelength waveguide arrays. We propose a subwavelength pitch MIR bent waveguide array with tapered Euler-shaped bends that are both complementary and asymmetric. In this design, the curvature of the Euler bends starts at 0 at the input and increases linearly, oscillating between 1/Rmin and -1/Rmin. The angle of the waveguide relative to its initial direction alternates between $\theta$ and -$\theta$, with these angles corresponding to the points of zero curvature in the bends \cite{pol1, pol2, pol3, pol4, pol5, pol6}.  We utilize Euler bends with tapered, asymmetric, and complementary widths, achieving crosstalk reduction below 30 dB for both the first and second neighboring waveguides across a \SI{500}{\nano\meter}  wavelength range ( \SIrange{3.1}{3.6}{\micro\meter}), resulting in an efficient OPA with a large field of view. Although this work is demonstrated on a silicon-on-insulator platform, the design principles can be extended to other material platforms such as indium phosphide (InP), gallium arsenide (GaAs), and germanium-on-silicon (Ge-on-Si). This adaptability allows the approach to cover mid-infrared wavelengths beyond \SI{3.6}{\micro\meter}. In all previously reported waveguide arrays, the waveguide widths are kept constant along the channels. Our design is the first to employ asymmetric and complementary tapered waveguide widths, providing greater fabrication flexibility while maintaining low crosstalk. Unlike prior approaches, it does not rely on strictly constant waveguide widths to achieve crosstalk suppression below \SI{30}{dB} over a broad wavelength range. Table~1 presents a comparison of this work with the literature on subwavelength-pitch waveguide arrays.

\begin{table}[htbp]
    \centering
    \caption{Comparison of this work with the previous work in literature for subwavelength pitch waveguide arrays}
    \label{tab:comparison}
    \begin{tabular}{lccccc}
        \toprule
        Ref. & Wavelength ($\mu$m) & Crosstalk & Bandwidth & Waveguide & Structure \\
        \midrule
        \cite{song2015high}       & 1.50 - 1.58 & $<20$ dB & 100 nm & straight & different widths of neighboring WGs \\
        \cite{shen2016increasing} & 1.52 - 1.58 & $<15$ dB & 60 nm & straight & nanophotonic cloaking  \\
        \cite{chen2022realization}& 1.50 - 1.56 & $<20$ dB & 60 nm & straight & deep-subwavelength silicon strips \\
        \cite{xu2016ultra}        & 1.52 - 1.62 & $<30$ dB & 100 nm  & bent & $90^o$  bend with constant WG widths\\
        \cite{zhou2023artificial} & 1.48 - 1.55 & $<30$ dB & 70 nm & bent & Artificial Gauge Field \\
        \cite{li2024morphology}   & 3.5 - 3.6  & $<30$ dB & 40 nm  & bent & Artificial Gauge Field \\
       \small{This work}  & 3.1 - 3.6   & $<30$ dB & 500 nm & bent & Complementary and Asymmetric Tapered WGs \\
        \bottomrule
    \end{tabular}
\end{table}

\section*{Numerical Methods and Design}

\begin{figure}[!b]
 	\centering
 	\begin{center}
 		
		\includegraphics[width=1\linewidth]{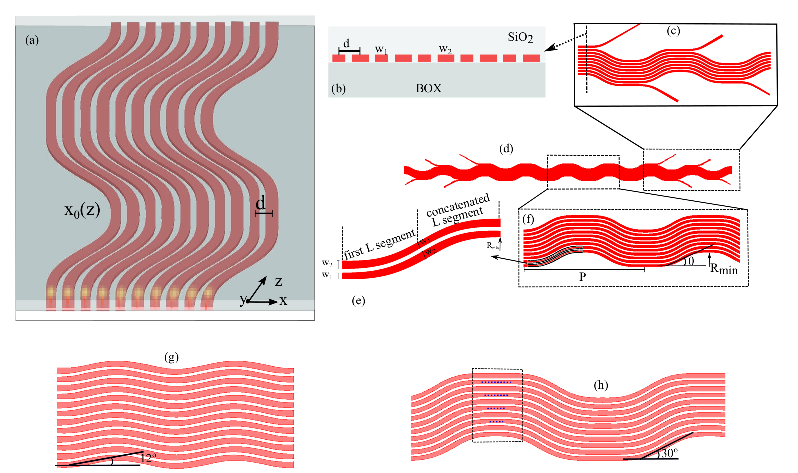}
				\caption{: Schematic illustration of the proposed waveguide array: (a) 3D view, (b) cross-sectional view, (c) enlarged view of the output section of the waveguide array, (d) \SI{200}{\micro\meter} long waveguide array with bend radius of \SI{12}{\micro\meter} and $30^o$ bend angle, (e) enlarged view of neighboring waveguides indicating complementary widths, (f) enlarged view of the middle section, (g) array with $12^o$ bend angle, (h) array with $30^o$ bend angle. }
		\label{schematicA}
		
	\end{center} 
\end{figure}

The 3D view and top view of the proposed PSR are shown schematically in Fig.~\ref{schematicA}(a) and Fig.~\ref{schematicA}(d) respectively. 
In this MIR bent waveguide array, the term 'asymmetric' refers to the design of the waveguide, in which the width is not constant. Instead, each L-shaped segment of the Euler bend features tapered widths: for example, the first L-shaped segment starts at width $w_1$~=~\SI{1}{\micro\meter} and gradually increases to $w_2$~=~\SI{1.3}{\micro\meter}. In contrast, the second concatenated L-shaped segment begins at \SI{1.3}{\micro\meter} and tapers back down to \SI{1}{\micro\meter}. The term "complementary" describes how these waveguides interact when arranged in an array. Neighboring waveguides are designed to complement each other in width. For instance, while one L-shaped segment starts at \SI{1.3}{\micro\meter}, the adjacent waveguide follows a pattern that ensures complementary widths (see Fig.~\ref{schematicA}(e)). This arrangement enhances the overall performance of the array by optimizing light propagation and minimizing crosstalk.
Fig.~\ref{schematicA}(a) shows a 3D view of the waveguide array with the bent trajectory $x_0(z)$, where light propagates in the z-direction. The equivalent propagation constant of the multi-waveguide model is given by the following formula: 
\begin{equation}
\beta_{\text{peq}} = \beta_0 + \kappa_0 + 2 \sum_{m > 0} \cos(k_x m d) \, \kappa_{m\text{eq}}
\end{equation}

here $\kappa_0$ and $k_x$ correspond to self-coupling coefficient and momentum of the relevant Bloch state, respectively, $\beta_0$ is the propagation constant of an individual straight waveguide; d is the distance between the center-to-center waveguide; $\kappa_{meq}$ is $\kappa_m e_m$ where $\kappa_m$ is the coupling coefficient to the mth nearest waveguide, and $e_m$ is the coupling coefficient variation factor, averaged over a full period and can be calculated as follows:   
\begin{equation}
e_m = \frac{1}{P} \int_0^P \exp\left[i k_0 n_s m d \frac{\partial x_0(z)}{\partial z} \right] \, dz.
\end{equation}

Here $n_s$ and $k_o$ are the mode effective index and wavenumber, respectively, and $\frac{\partial x_0(z)}{\partial z}$ is the derivative of the curved waveguide path. In the present design, the waveguide path is chosen to follow an Euler curve. This allows for defining an array of waveguides with Euler-based bends that vary in bend angles and radii. Fig.~\ref{schematicA}(d) illustrates an array of waveguides with a bend angle of $30^o$ and a minimum radius of \SI{12}{\micro\meter}. Fig.~\ref{schematicA}(f) provides an enlarged view, while Fig.~\ref{schematicA}(e) shows a close-up of two adjacent waveguides.
In Fig.~\ref{schematicA}(e), it is evident that the width of the lower waveguide starts at $w_1$~=~\SI{1}{\micro\meter}, reaches a maximum width of $w_2$~=~\SI{1.3}{\micro\meter} at the end of the first L-shaped segment, and then reduces back to $w_1$~=~\SI{1}{\micro\meter} at the end of the concatenated L segment. Conversely, the width of the adjacent upper waveguide starts at the wider $w_2$~=~\SI{1.3}{\micro\meter}, narrows to $w_1$~=~\SI{1}{\micro\meter} in the middle, and then expands back to $w_2$ at the end. This is because, along the propagation path, the two waveguides maintain dissimilar widths and therefore support modes with different effective indices, which suppresses coupling. Additionally, the changing curvature in light propagation further reduces crosstalk by controlling the complex phase during coupling. The waveguides are arranged at a subwavelength scale, with a center-to-center distance of half the operating wavelength. This arrangement leads to an extremely low crosstalk, below 30dB in the first and second neighboring waveguides, which cannot be achieved so far with such a low crosstalk in both neighbors at a broad wavelength range of 500nm (\SIrange{3.1}{3.6}{\micro\meter} wavelength range). 
\\The two waveguides in Fig.~\ref{schematicA}(e) are connected back-to-back to form a curved trajectory along the z-axis, as shown in Fig.~\ref{schematicA}(a) (indicating 10 channels). Extending this two-waveguide configuration along the x-axis, as in Fig.~\ref{schematicA}(f), creates a 10-channel waveguide array. For smaller bend angles, the number of channels can be scaled up significantly, potentially to hundreds, because the radius of each waveguide remains nearly constant, resulting in minimal variation in crosstalk between channels. This can be seen in Fig.~\ref{schematicA}(g), where a 12-degree bend angle allows all channels to maintain the same bend radius. However, with larger bend angles, the sharper curvature results in a change in radius for each added waveguide channel, leading to variations in crosstalk. To maintain extremely low crosstalk and consistent waveguide radius across channels, adjustments are needed. For example, in Fig.~\ref{schematicA}(h), the 30-degree bends require modifications to preserve the waveguide radius as more channels are added. Here, the S-shaped sections of the lowest waveguide connect directly. Moving upward in the array, each additional waveguide channel requires small straight segments at the connection points between S-bends to maintain spacing and radius, as highlighted by blue dotted lines in Fig.~\ref{schematicA}(h). These straight sections do not increase crosstalk due to the phase mismatch created by the differing waveguide widths (1 and 1.3 microns) in adjacent waveguides. The cross-section of this waveguide array and the output section of the waveguide array are shown in Fig.~\ref{schematicA}(b) and Fig.~\ref{schematicA}(c), respectively.
\begin{figure}[!ht]
 	\centering
 	\begin{center}
 		
		\includegraphics[width=0.8\linewidth]{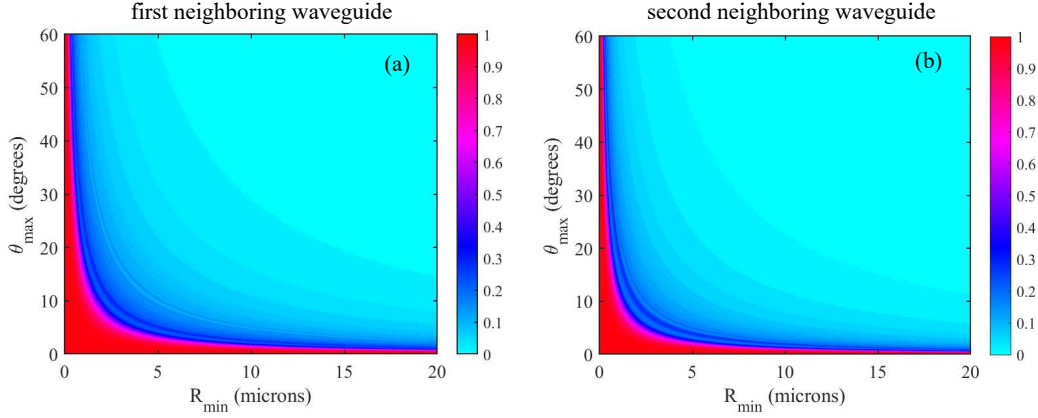}
				\caption{The coupling coefficient variation factor as a function of bend angle and radius for the \SI{3.55}{\micro\meter} wavelength and center-to-center spacing of \SI{1.775}{\micro\meter}, (a) first neighboring waveguide, (b) second neighboring waveguide }
		\label{fig2}
		
	\end{center} 
\end{figure}

The waveguide array is defined on a silicon-on-insulator (SOI) platform, with a 2-micron buried oxide (BOX) layer and covered with $ \mathrm{SiO}_2 $ from the top.  The bends are defined as adiabatic curves on a 500nm-thick silicon layer, where the curvature increases linearly with their length.  To determine the coupling strengths to neighboring waveguides, we calculate the coupling coefficient variation factors for the first two neighboring waveguides. Crosstalk with the third neighboring waveguide can be neglected due to the exponential decrease in the coupling coefficient with increasing waveguide separation. To calculate the coupling coefficient variation factor, we first find $\frac{\partial x_0(z)}{\partial z}$  of the waveguide trajectory, with the parametric equations for the L-shaped Euler path with the path length $L = 2R\theta$  provided below:
\begin{equation}
\begin{split}
x &= \int_0^L \cos(s^2) \, ds, \\
z &= \int_0^L \sin(s^2) \, ds
\end{split}
\end{equation}

The analytical solution of Equation 2 is calculated and plotted in Fig.~\ref{fig2} for the first and second neighboring waveguides for the \SI{3.55}{\micro\meter} wavelength and at subwavelength pitch with center-to-center spacing of \SI{1.775}{\micro\meter}. Fig.~\ref{fig2}(a) and Fig.~\ref{fig2}(b) depict the coupling coefficient variation factor as a function of minimum bend radius and maximum angle for the first and second neighboring waveguides, respectively. They indicate an extremely low variation factor, reaching zero within the radius range of 5 to 50 microns and the angle range of 25 to 45 degrees. This further suggests a significant flexibility in adjusting the bend size according to the area requirements.

\section*{Results discussion and fabrication tolerance}
The spectral responses of the waveguide array were simulated using the 3D finite-difference time-domain (FDTD) method. 
\begin{figure}[!ht]
 	\centering
 	\begin{center}
 		
		\includegraphics[width=0.9\linewidth]{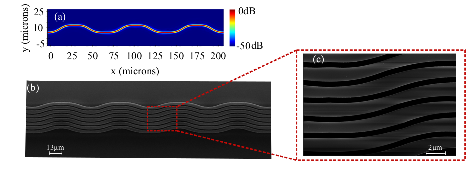}
				\caption{(a) top view illustrates the propagation of light at \SI{3.55}{\micro\meter} wavelength, showing the electric field intensity of the TE mode entering from the middle channel of a 10-channel waveguide array, indicating negligible light coupling to the neighboring waveguides. (b) SEM images of the waveguide arrays fabricated using electron beam lithography}
		\label{fig3}
		
	\end{center} 
\end{figure}
These responses were calculated by launching a transverse electric (TE) mode into each input of a 10-channel waveguide array, \SI{200}{\micro\meter} in length, with a \SI{12}{\micro\meter} radius and a $30^o$ bend angle.
\begin{figure}[!ht]
 	\centering
 	\begin{center}
 		
		\includegraphics[width=1\linewidth]{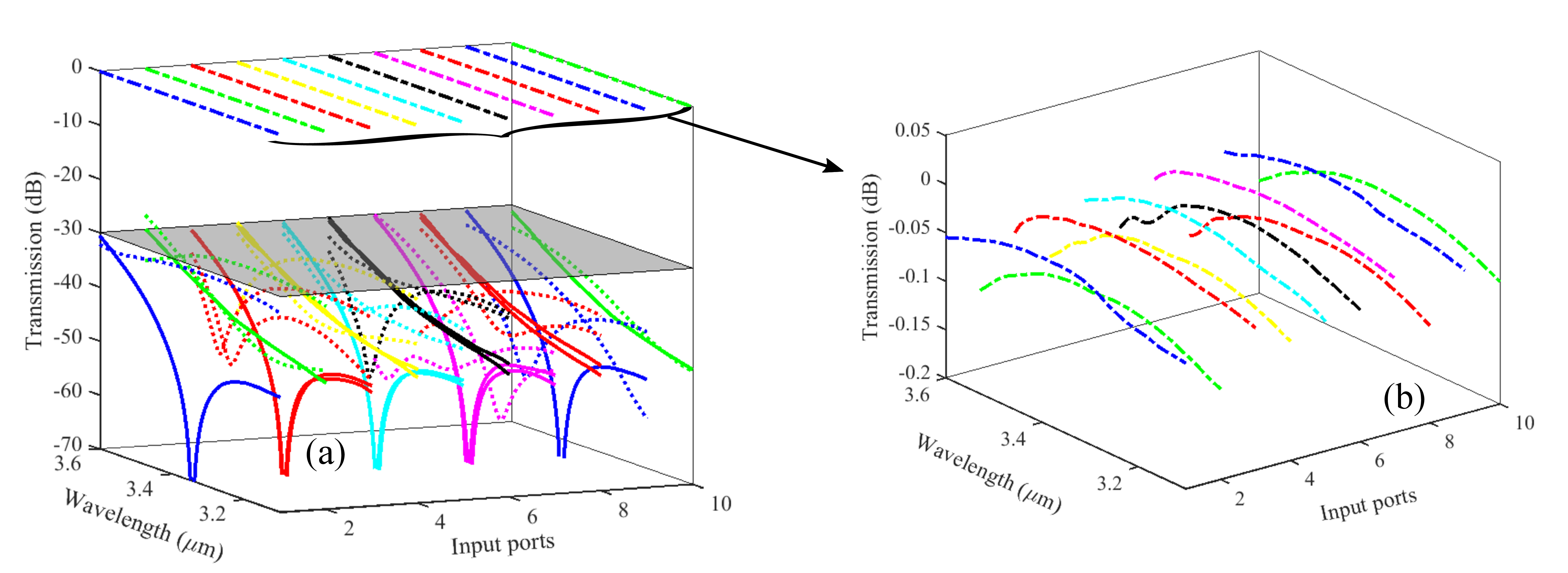}
				\caption{(a) The simulated spectral transmissions for a 10-channel waveguide array are shown, with dashed lines indicating the insertion loss, and dotted and solid lines representing the crosstalk in the first and second neighboring channels on both sides of the input waveguide, respectively. (b) Enlarged view of the insertion loss.}
		\label{fig4}
		
	\end{center} 
\end{figure}
As shown in Fig.~\ref{fig3}(a), the desired negligible coupling is evident in the electric field distribution of the TE mode propagating through the central channel of the array. At a wavelength of \SI{3.55}{\micro\meter}, light propagation through the \SI{200}{\micro\meter}-long array exhibits minimal coupling to adjacent waveguides. The spectral responses were simulated over a broad wavelength range of \SIrange{3.1}{3.6}{\micro\meter}, as illustrated in Fig.~\ref{fig4}.
The transmissions are plotted for the input channel and the two neighboring channels on either side. The dashed lines in Fig.~\ref{fig4}(a) represent the transmission through the input channel (i.e., the insertion loss), with an enlarged view shown in Fig.~\ref{fig4}(b), indicating an extremely low insertion loss of below 0.1 dB over the wavelength range of \SIrange{3.1}{3.6}{\micro\meter}. On the other hand, the dotted and solid lines in  Fig.~\ref{fig4}(a) correspond to the crosstalk in the first and second neighboring waveguides, respectively, on both sides of the input channel. It can be seen that excellent broadband performance, with extremely low crosstalk below -30 dB in both neighboring channels, has been achieved over the wavelength range of \SIrange{3.1}{3.6}{\micro\meter}.

The proposed waveguide array is highly fabrication-tolerant because the waveguide widths vary while maintaining a constant center-to-center spacing. This prevents changes in coupling strength, even with fabrication-induced width variations, making the design robust and fabrication-friendly. As mentioned earlier, we design an L-shaped Euler bend with a tapered width profile. The first bend begins with a width of \SI{1}{\micro\meter} and gradually increases to \SI{1.3}{\micro\meter} at the output. A complementary bend is placed adjacent to it, starting at \SI{1.3}{\micro\meter} and tapering down to \SI{1}{\micro\meter}. This tapered geometry improves fabrication tolerance. Since the waveguide width varies continuously over a relatively large range of 300 nm, small fabrication deviations, such as ±10 nm or even larger, are minor in comparison and therefore have little impact on device performance. Likewise, small variations caused by sidewall angle imperfections are effectively mitigated by the taper. As a result, the proposed structure offers both low crosstalk and high robustness against fabrication-induced imperfections. Furthermore, the waveguide arrays are optimized for the TE mode, which is usually well confined inside the core. Its field intensity at the sidewalls is lower compared to TM. This makes it less sensitive to sidewall roughness. Additionally, the waveguides are not made of tiny slots, which makes fabrication flexible. The design is compatible with standard CMOS fabrication and can be realized using a single etch-step lithography process. The design is highly amenable to fabrication and can be realized with precise dimensional accuracy utilizing advanced optical \cite{zafar2019compact} and electron-beam lithography techniques \cite{zafar2020low}. The proof-of-concept design was fabricated with 7 channels, as shown in Fig.~\ref{fig3}(b), demonstrating the high accuracy of the waveguide dimensions. The waveguide arrays were fabricated commercially by Applied Nanotools Inc. using a proprietary process. This process is based on Inductively Coupled Plasma (ICP) reactive ion etching and produces highly vertical sidewalls. As discussed above, even if small sidewall angle deviations are present, the tapered profile ensures that device performance is not adversely affected. The same fabrication facility was previously used to realize waveguide arrays reported in \cite{zafar2023band, zafar2023compact}, which were optimized for telecommunication bands in the \SIrange{1.3}{1.7}{\micro\meter} wavelength range. Measurements of those devices, as presented in the cited works, confirmed negligible insertion loss and ultra-low crosstalk, demonstrating that small variations in sidewall roughness and angle inherent to this fabrication process do not introduce significant losses. The array is implemented on a silicon layer with a thickness of \SI{500}{\nano\meter}, where the typical variation across an entire wafer is approximately ±10 nm, with a 3-sigma uncertainty. The structures were patterned using electron beam lithography, and covered with a \SI{2}{\micro\meter}-thick oxide overcladding layer.

\section*{Conclusion}
In conclusion, the proposed mid-infrared waveguide array design offers a significant advancement in minimizing crosstalk and achieving subwavelength-pitch integration. By employing complementary and asymmetric tapered Euler-shaped bends, the design achieves crosstalk reduction below 30 dB across a broad wavelength range: from \SIrange{3.1}{3.6}{\micro\meter}.  The innovative use of asymmetric and complementary waveguide widths optimizes light propagation and enhances the overall performance of the array. The design is highly fabrication-tolerant, maintaining consistent performance despite potential width variations, and is compatible with standard CMOS fabrication processes.  The results of this study pave the way for more efficient and compact optical phased arrays, with applications in various fields such as LiDAR systems, free-space communications, and integrated photonic circuits.

\section*{Author contributions statement}

M.F.P. and H.Z. conceived the study. H.Z. performed the numerical simulations.  M.F.P. supervised the research and secured funding. Both authors analysed the results and reviewed the manuscript.

\section*{Funding}
This publication is based upon work supported by Khalifa University under Award No. CIRA-2021-108

\section*{Competing interests} The authors declare no competing interests. 

\section*{Data availability} The data used to generate the figures and tables can be obtained from the corresponding author upon reasonable request.
\bibstyle{nature}
\bibliography{sample}

\end{document}